\documentclass[preprint,tighten,eqsecnum,floats,epsfig,aps]{revtex4}
\usepackage{graphicx}
\begin{document}
\def\mbphi{\mbox{\boldmath$\phi$}}
\def\sq{\sqrt{2}}
\def\bit{\begin{itemize}}
\def\eit{\end{itemize}}
\def\bnu{\begin{enumerate}}
\def\enu{\end{enumerate}}
\def\sss{\scriptscriptstyle}
\def\ss{\scriptstyle}
\def\rh {\hat{r}}
\def\fb {{\bf f}}
\def\Sb{ {\bf S}}
\def\xb {{\bf x}}
\def\Xb {{\bf X}}
\def\A {{{\cal A}}}
\def\B {{{\cal B}}}
\def\Q {{{\cal Q}}}
\def\R {{{\cal R}}}
\def\Qb{ {\bf Q}}
\def\nn{\nonumber }
\def\qb {{\bf q}}
\def\mbt{\mbox{\boldmath$\tau$}}
\def\mbr{\mbox{\boldmath$\rho$}}
\def\bin#1#2{\left(\negthinspace\begin{array}{c}#1\\#2\end{array}\right)}
\def\mbp{\mbox{\boldmath$\phi$}}
\def\mbxi{\mbox{\boldmath$\xi$}}
\def\mbpi{\mbox{\boldmath$\pi$}}
\def\boxp{\mbox{\boldmath$p$}}
\def\boxq{\mbox{\boldmath$q$}}
\def\mbs{\mbox{\boldmath$\sigma$}}
\def\mbn{\mbox{\boldmath$\nabla$}}
\def\H {{{\cal H}}}
\def\M {{{\cal M}}}
\def\x{\times}
\def\Ket#1{||#1 \rangle}
\def\Bra#1{\langle #1||}
\def\lsim{\:\raisebox{-0.5ex}{$\stackrel{\textstyle<}{\sim}$}\:}
\def\gsim{\:\raisebox{-0.5ex}{$\stackrel{\textstyle>}{\sim}$}\:}
\def\ie{{\em i.e., }}
\def\nn{\nonumber }
\def\be{\begin{equation}}
\def\ee{\end{equation}}
\def\br{\begin{eqnarray}}
\def\er{\end{eqnarray}}
\def\brn{\begin{eqnarray*}}
\def\ern{\end{eqnarray*}}
\def\etc{ {\it etc}}
\def\qb {{\bf q}}
\def\pb {{\bf p}}
\def\Pb{ {\bf P}}
\def\rb {{\bf r}}
\def\Rb{ {\bf R}}
\def\e {{\epsilon}}
\def\mbl{\mbox{\boldmath$\lambda$}}
\def\mbs{\mbox{\boldmath$\sigma$}}
\def\mbt{\mbox{\boldmath$\tau$}}
\def\bra#1{\langle #1|}
\def\ket#1{|#1 \rangle}
\def\rf#1{{(\ref{#1})}}
\def\ov#1#2{\langle #1 | #2  \rangle }
\def\sixj#1#2#3#4#5#6{\left\{\negthinspace\begin{array}{ccc}
#1&#2&#3\\#4&#5&#6\end{array}\right\}}
\def\ninj#1#2#3#4#5#6#7#8#9{\left\{\negthinspace\begin{array}{ccc}
#1&#2&#3\\#4&#5&#6\\#7&#8&#9\end{array}\right\}}
\def\ss{\scriptstyle}
\def\go{\rightarrow  }
\def\etal{{\it et al. }}
\def\E {{{\cal E}}}
\def\I {{{\cal I}}}
\def\V {{{\cal V}}}
\def\sqi{\frac{1}{\sqrt{2}}}
\def\fot{\frac{1}{2}}


\title{Nuclear Structure in Nonmesonic Weak Decay of Hypernuclei}

\author{ F. Krmpoti\'c}

\affiliation{Departamento de F\'{\i}sica, Universidad Nacional de
La Plata, C. C. 67, 1900 La Plata, Argentina}

\author{  D. Tadi\'c}

\affiliation{Physics Department,University of Zagreb, 10\,000 Zagreb, Croatia}

\date{\today}
\begin{abstract}
A general shell model formalism for the nonmesonic weak decay of the
hypernuclei has been developed.
It involves a partial wave expansion of the emitted nucleon waves,
preserves   naturally  the antisymmetrization between the
escaping particles and the residual  core, and contains as a particular case
the weak $\Lambda$-core coupling formalism.
The hypernuclei are grouped having in view their $A-1$ cores, that  is in 
those with even-even, even-odd and odd-odd cores.
It is shown  that in all three cases the  nuclear structure
manifests itself basically through  Pauli Principle, and  very simple 
expressions are derived  for the neutron and proton induced decays
rates,  $\Gamma_n$ and $\Gamma_p$, which does not involve the
spectroscopic factors.
For the  strangeness-changing weak $\Lambda N\to NN$ transition
potential we use the One-Meson-Exchange Model (OMEM), which
comprises  the exchange of the complete pseudoscalar and vector meson octets
($\pi,\eta,K,\rho,\omega,K^*$). We evaluate ${^{3} H}$, ${^{4} H}$, 
${^{4}_\Lambda He}$, ${^{5}_\Lambda He}$,${^{11} B}$,
${^{12}_\Lambda C}$, ${^{16}_\Lambda O}$, ${^{17}_\Lambda O}$,
and ${^{28}_\Lambda Si}$ hypernuclei, with  commonly used
parametrization for  the OMEM,
and compare the results with the available experimental information.
The calculated  rates $\Gamma_{NM}= \Gamma_n+\Gamma_p$ are
consistent with the data, but the measurements  of  
$\Gamma_{n/p}=\Gamma_n/\Gamma_p$ are not well accounted for by the
 theory.
It is suggested that, unless additional degrees of freedom are incorporated,
the OMEM parameters should be radically modified.
\end{abstract}
\maketitle

\section{Introduction}
Hypernuclear physics adds another flavor (strangeness) to the traditional nuclear physics, and
its goal is to study the behavior of  hyperons ($\Lambda,\Sigma,\Xi,\Omega)$ in the nuclear
environments, which are now bound system of neutrons, protons and one or more hyperons.
Interesting strange nuclei with strangeness $S=-1$ are the $\Lambda$ hypernuclei, in which a
$\Lambda$ hyperon, having a mass of $1116$ MeV and zero charge and isospin, replaces
one of the nucleons.
Same as the  free $\Lambda$ hyperon, they
are mostly produced via the strong interactions, \ie in the
reaction processes $\pi^+n\go\Lambda K^+$, $K^-n\go\pi^-\Lambda$ and
$K^-p\go\pi^0\Lambda$, by making use of the pion ($\pi$) and kaon ($K$) beams.
They also
 basically decay  through
the weak interactions, as the free $\Lambda$ does. Yet, as it is well known
and explained below,
there are some very important differences in the corresponding decaying modes.

First, it should be remembered that the  free $\Lambda$ hyperon decays
nearly $100$ \% of the time by the $\Lambda\go N\pi$ weak-mesonic mode
(Fig. 1):
\vspace{1cm}
\begin{figure}[h]
\centerline{\includegraphics[scale=0.6]{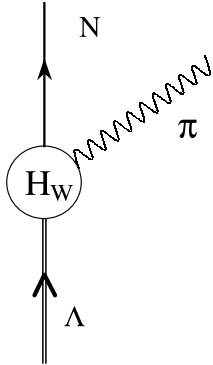}}
\caption{ Mesonic (nonleptonic) decay vertex $\H^W_{\Lambda N\pi }$. \label{fig1}}
\end{figure}

\brn
 \Lambda\go \left\{\begin{array}{ccc}
 p+\pi^{-} && (64.1\%) \\
         n+\pi^{0} && (35.7\%),
\end{array}\right.
\ern
with the total transition rate $\Gamma^0_{\pi^-}+\Gamma^0_{\pi^0}=\Gamma^0= 2.50
\cdot 10^{-6}$ eV (which corresponds to the lifetime $\tau^0= 2.63
\cdot 10^{-10}$ sec). For the decay at rest  the energy-momentum conservation implies
\brn
M_\Lambda=M_N+\frac{p_N^{2}}{2M_N}+\sqrt{p_\pi^{2}+m_\pi ^{2}};~~~~p_N&\equiv&p_\pi.
\ern
Therefore the energy released is
\brn
Q_0=M_\Lambda-M_N-m_\pi\cong 37~ \mbox{MeV},
\ern
and the kinetic energies and momenta in the final state are:
\brn
T_N&=&\frac{(M_\Lambda-M_N) ^{2}+m_\pi ^{2}}{2M_\Lambda}\cong 5~ \mbox{MeV};
~~~T_\pi=Q_0-T_N\cong 32~ \mbox{MeV},\nn\\
p_N&\equiv&p_\pi=\sqrt{(T_N+M_N)^2-M_N^2}\cong 100~ \mbox{MeV}/c.
\ern
 It is clear that
  the isospin is changed by $\Delta T=1/2$ and $3/2$
and its projection by $\Delta M_T=-1/2$  in the free $\Lambda\go N\pi$ decay. More, as
\bit
\item{$ \Lambda \go n+\pi^0$:}
\brn
&& \ket{n}\ket{\pi^0}\equiv \ket{1/2,-1/2}\ket{10}
=\sqrt{\frac{1}{3}}\ket{1/2,1;1/2,-1/2}\
+\sqrt{\frac{2}{3}}\ket{1/2,1;3/2,-1/2}\
\nn\\\ern
\item{$\Lambda \go p+\pi^-$:}
\brn
&& \ket{p}\ket{\pi^-}\equiv \ket{1/2,1/2,}\ket{1,-1}
=-\sqrt{\frac{2}{3}}\ket{1/2,1;1/2,-1/2}\
+\sqrt{\frac{1}{3}}\ket{1/2,1;3/2,-1/2},\
\nn\\\ern
\eit
one sees that the above experimental data can be accounted for fairly well by
neglecting the $\ket{1/2,1;3/2,-1/2}$ components in these relations ($\Delta T=1/2$ rule).
In fact, one gets $\Gamma_{\pi^-}/\Gamma_{\pi^0}=2$, while the experimental result is
$64.1/35.7=1.80$.

Assuming the $\Delta T=1/2$ rule, the phenomelogical weak  Hamiltonian for the
process depicted in Fig. 1 can be expressed as:
\br
\H^W_{\Lambda N\pi }&=&-iG_Fm_\pi^2\overline{\psi}_N\left(A_\pi+B_\pi\gamma_5\right)
\mbphi_\pi\cdot\mbt\psi_\Lambda\bin{0}{1},
\label{1.1}\er
where $G_Fm_\pi^2=2.21\x 10^{-7}$ is the weak coupling constant. The empirical constants
$A_\pi=1.05$  and $B_\pi=-7.15$, adjusted to the observables of the free $\Lambda$ decay,
determine the strengths of parity violating and parity conserving amplitudes, respectively.
The nucleon, $\Lambda$ and pion fields are given by $\psi_N$ and  $\psi_\Lambda$
and $\mbphi_\pi$, respectively, while the isospin spurion $\bin{0}{1}$ is included in order
to enforce the empirical $\Delta T={ 1/2}$ rule.
(Note that: $\overline{\psi}_N\mbphi_\pi\cdot\mbt\psi_\Lambda=
\left(\overline{\psi}_{p}\phi_{\pi^0}+\sq\overline{\psi}_n\phi_{\pi^-},
\sq\overline{\psi}_{p}\phi_{\pi^+}-\overline{\psi}_n\phi_{\pi^0}\right)\Lambda$).

 The free $\Lambda$ hyperon weak decay  is radically modified  in the nuclear
environment because the nucleon and  the hyperon now move, respectively,
  in  the mean fields $U_N$ and $U_\Lambda$, which come from the $NN$ and $N\Lambda$
  interactions. $U_N$ and $U_\Lambda$ are characterized by the single particle energies
(s.p.e.)  $\varepsilon_N$ and $\varepsilon_\Lambda$
and we have to differentiate between:

1. {\it Mesonic Decay (MD)}:
 The basic process is again represented by the graph shown in Fig. 1 and
  described by the hamiltonian \rf{1.1}. Yet,
the energy-momentum conservation is different:
\brn
M_\Lambda=M_N-\varepsilon_\Lambda
+\varepsilon_N^\uparrow +\frac{p_A^{2}}{2M_A}+\sqrt{p_\pi^{2}+m_\pi ^{2}};~~~~\pb_A&=&-\pb_\pi
\ern
where $M_A=AM_N$ and $\pb_A$ are, respectively,  the mass and the momentum
of the whole nucleus; $A$ is the mass number, and $\varepsilon_N^\uparrow$
are the s.p.e. of the loosely bound states above the Fermi energy $\varepsilon_{N}^{F}$.
They are of the order of a few MeV, while
$\varepsilon_\Lambda$ is the energy of  the $0s_{1/2}$ state and goes from $-11.7$~MeV for
${^{13}_\Lambda C}$ to $-26.5$~MeV for ${^{208}_\Lambda Pb}$ \cite{Us99}.
Thus, the corresponding Q-values
\brn
Q_M=M_\Lambda-M_N-m_\pi+\varepsilon_\Lambda
-\varepsilon_N^\uparrow,
\ern
are significantly smaller than
$Q_0$, particularly for medium and heavy nuclei.
The experimental decay rates
$\Gamma_{\pi^-}+\Gamma_{\pi^0}=\Gamma_M\equiv \Gamma_M 
(\Lambda\rightarrow N\pi$)
 are of the order of
$\Gamma^0$ only for  nuclei with $A\le 4$, and they rapidly fall
as a function of nuclear mass. For instance, in ${^{12}_\Lambda C}$:
$\Gamma_{\pi^0}/\Gamma^0=0.217\pm 0.084$ and
$\Gamma_{\pi^-}/\Gamma^0=0.052^{+0.063}_{-0.035}$.
This hindrance effect, as is illustrated in Fig. 2 for the hypernucleus 
${^{17}_\Lambda O}$, is  due to Pauli principle.
In fact, the population of states that are below the Fermi level, with energies
$\varepsilon_N^\downarrow\le\varepsilon_N^F$, is totally blocked by
 the Pauli principle, while transitions  to  states that lie above
$\varepsilon_{N}^{F}$ are strongly hindered due to the selection rule 
$\Delta$N$=0$, N being  the harmonic oscillator quantum numbers. That is,
only a few  second forbidden transitions ($\Delta$N$=2$) can occur in
 the case of ${^{17}_\Lambda O}$, and it is clear that the degree of 
forbiddiness increases with $A$.

\begin{figure}[h]
\centerline{\includegraphics[scale=0.5]{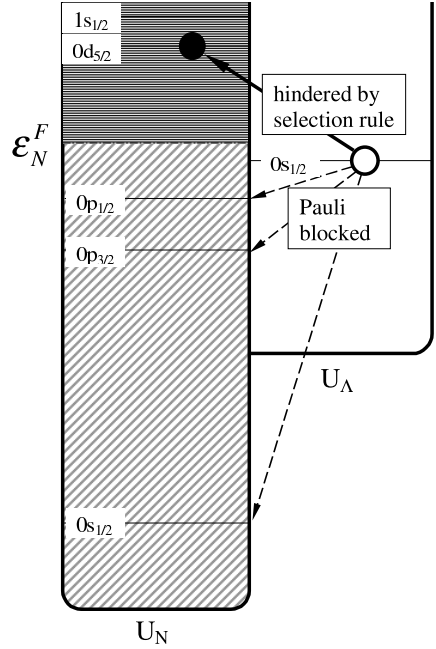}}
\caption{ Schematic picture of the hypernuclear mesonic decay in $^{17}_\Lambda O$:
The population of occupied states $0s_{1/2},0p_{3/2}$ and $0p_{1/2}$, which are below the Fermi
level $\varepsilon_N^F$, is totally blocked by the Pauli principle,
while transitions  to  the weakly bound empty states $0d_{5/2}$ and $1s_{1/2}$, which
lie above $\varepsilon_{N}^{F}$, are strongly hindered by the selection rule $\Delta$N$=0$.\label{fig2}}
\end{figure}

2. {\it Nonmesonic Decay}:
New nonmesonic  decay (NMD) channels $\Lambda N\rightarrow NN$
 become open inside the nucleus, where there are no
pions in the final state.  The
corresponding transition rates can be stimulated either by
protons, $\Gamma_p\equiv\Gamma(\Lambda p\rightarrow np)$, or by
neutrons, $\Gamma_n\equiv\Gamma(\Lambda n\rightarrow nn)$.
The energy-momentum conservation and the Q-value are, respectively:
\brn
M_\Lambda&=&M_N-\varepsilon_\Lambda
-\varepsilon_N^\downarrow +\frac{p_1^{2}}{M_N}+\frac{p_2^{2}}{M_N}+\frac{p_A^{2}}{AM_N}
;~~~~\pb_A=-\pb_1-\pb_2,
\ern
and
\brn
Q_{NM}=M_\Lambda-M_N+\varepsilon_\Lambda
+\varepsilon_N^\downarrow,
\ern
where $\pb_1$ and $\pb_2$ are the momenta
of the two outgoing nucleons. As the mean value of
$\varepsilon_N^\downarrow$ is $\sim 30$ MeV
one gets that $Q_{NM}\sim 120-135$ MeV, which is basically the kinetic energy
of the two  particles that are ejected from the hypernucleus.
This means that the nonnesonic decay process possesses a large phase
space in the continuum, as  is outlined  in Fig. 2 for the case of the 
hypernucleus ${^{17}_\Lambda O}$.
The theoretical models reproduce fairly
well the experimental values of the  total width
$\Gamma_{NM}=\Gamma_n+\Gamma_p$ ($\Gamma_{NM}^{\rm exp}\cong \Gamma^0$)
 but the ratio $\Gamma_{n/p}\equiv\Gamma_n/\Gamma_p$ ($0.5\le \Gamma_{n/p}^{\rm
exp}\le 2$) remains a puzzle.

\begin{figure}[h]
\centerline{\includegraphics[scale=0.5]{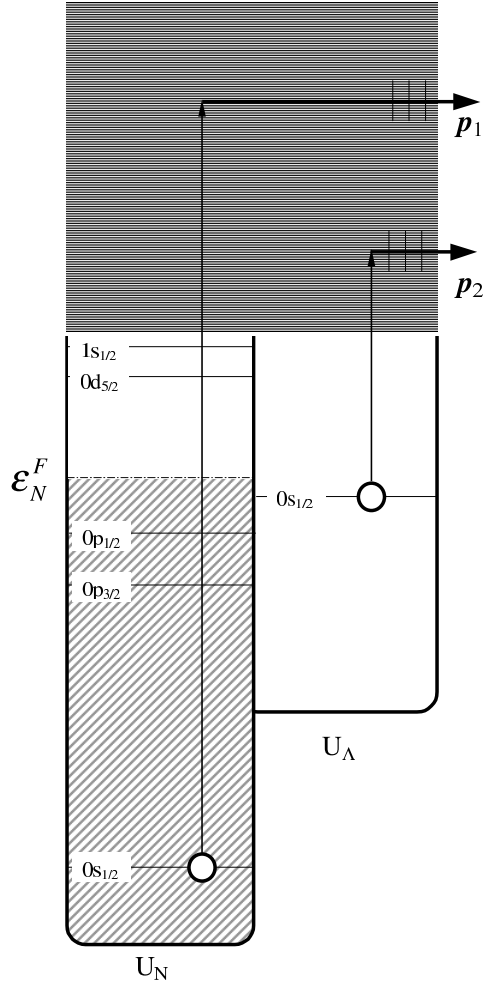}}
\caption{ Schematic picture of the hypernuclear nonmesonic decay in 
$^{17}_\Lambda O$:
The hyperon $\Lambda$ and one of the nucleons from the occupied  states
$0s_{1/2},0p_{3/2}$ and $0p_{1/2}$ are expelled into the large phase
space in the continuum,
becoming  free nucleons with momenta $\pb_1$ and $\pb_2$.
\label{fig3}}
\end{figure}

Very often  it is assumed that the  hypernuclear NMD
$\Lambda N\rightarrow NN$
 is triggered via the exchange of a virtual meson, and the obvious candidate
 is the one-pion-exchange (OPE) mechanism, where the
 strong  Hamiltonian
\br
\H^S_{N N\pi
}&=&ig_{NN\pi}\bar{\psi}_N\gamma_5\mbpi\cdot\mbt\psi_N,
\label{1.2}\er
(with $g_{NN\pi}=13.4$) accompanies the weak Hamiltonian \rf{1.1}.
Following the pioneering investigations of Adams \cite{Ad67}
several calculations  have been done within this coupling scheme
yielding: $\Gamma_{NM}^{\rm (OPE)}\cong {\Gamma^0}$ and
$\Gamma_{n/p}^{\rm (OPE)}\cong 0.1-0.2$
\cite{Mc84,Ta85,He86,Ba90,Co90,Al91,Ra92,Ra95,Pa95,Pa95a,Du96,Pa97,Ra97,Os98,Pa98,In98,It98,Pa99,Alb00,Al00,Sa00,Sa00a,Pa01,Os01,It02,Ba02}.
The importance of the $\rho$ meson in the weak decay mechanism
was first discussed by McKellar and Gibson \cite{Mc84}, and
the present-day consensus is, however, that
the effect of the $\rho$-meson on both $\Gamma_{NM}$ and
$\Gamma_{n/p}$ is small \cite{Pa95,Du96,Pa97,Os98,Ba02}.
The full  one meson-exchange  model (OMEM),
which encompasses  all pseudoscalar mesons ($\pi, \eta, K$) and
all vector mesons ($\rho, \omega, K^*$), has been also considered by several
authors  \cite{Du96,Pa97,Sa00,Sa00a,Pa01,Os01,Ba02}. From these works
we have learned that,
although the $K$ meson contribution  significantly  increases the
ratio $\Gamma_{n/p}$, the OMEM is unable to account for  the corresponding
experimental values.

\begin{figure}
\centerline{\includegraphics[scale=0.5]{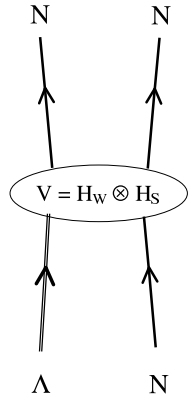}}
\caption{ The nonmesonic weak decay mode $\Lambda+N\go N+N$.
In the one meson exchange model a weak vertex $\H_W$
is always combined with a strong vertex $\H_S$. 
\label{fig4}}

\end{figure}

We wish to restate  that the  OMEM transition potential is
purely phenomenological and that it is not derived from a fundamental 
underlying form, as happens for instance, in the case of
electro-magnetic transitions or the semileptonic weak decays.
As in the case of the OPE, in the OMEM, a weak baryon-baryon-meson (BBM)
coupling is always combined with a strong BBM coupling (see Fig. 4). 
The  strong one
is determined experimentally with some help from the SU(3)
symmetry, and the involving   uncertainties have been copiously discussed in
the literature \cite{Na77,Ma89,Ha93,Pa94}. It is the weak BBM
couplings which could become  the largest source of errors. In
fact,  only the weak $N\Lambda\pi$ amplitude can be taken from the
experiment, at the expense of  neglecting the off-mass-shell
corrections. All other weak BBM couplings are derived theoretically by
using  $SU(3)$  and $SU(6)_w$ symmetries, octet
dominance, current algebra, PCAC, pole dominance, \etc.
\cite{Co90,Du96,Pa97,De80,Ma69,To82,Ok82,Co83,Na88,Ba01,Ba01a}.
Assortments of such methods have been developed and employed for a long time
in weak interaction physics to explain the hyperon nonleptonic decays.
One should also keep in mind that both the strong and weak BBM
couplings, as well the meson masses, can become significantly
renormalized by the nuclear environment \cite{Br02}.

The nuclear structure frameworks utilized in the literature for the
formal derivations of the NMD rates are: i) the nuclear matter, and 
ii) the nuclear shell model.
There are  relatively few works where the second method was  employed, and
up to quite recently all they were involved the technique of
coefficients of fractional parentage, with the spectroscopic factors
(SF) explicitly   appearing  in  the expressions for
the transition rates \cite{Ra92,Pa97,It98,It02}.
At variance, Barbero {\etal} \cite{Ba02}
have developed a fully general shell model formalism and
 have specified it for hypernuclei with odd-mass core, such as 
${^{4}_\Lambda H}$,  ${^{4}_\Lambda He}$, ${^{12}_\Lambda C}$, and 
${^{28}_\Lambda Si}$, were the cores are:
${^{3} H}$, ${^{3} He}$, ${^{11} C}$ and  ${^{27} Si}$. Here we also discuss
the hypernuclei
with even-even   and odd-odd number of protons and neutrons, namely
${^{5}_\Lambda He}$ and
 ${^{17}_\Lambda O}$,  and
${^{3}_\Lambda H}$, and
 ${^{11}_\Lambda B}$.
\newpage
\section{Shell Model Formalism}
The shell model framework for the NMD rates has been developed in
detail in Ref. \cite{Ba02} and here we will just sketch the main
steps, employing the same notation. One starts from Fermi's  Golden
Rule for the decay rate,
\br
\Gamma&=&2\pi \sum_{SM_SJ_FM_FTM_T} \int |\bra{\pb_1\pb_2 SM_S,J_FM_F;TM_T}
V\ket{J_IM_I}|^2
\nn\\
&\x&\delta(\epsilon_{p_1}+\epsilon_{p_2}+\E_F -\E_I)
\frac{d{\bf p}_1}{(2\pi)^3}\frac{d{\bf p}_2}{(2\pi)^3},
\label{2.1}\er
where  $V$ is  the weak hypernuclear potential,
and  the wave functions for the kets $\ket{\pb_1\pb_2 SM_S,J_FM_F;TM_T}$
and $\ket{J_IM_I}$ are assumed to be antisymmetrized  and normalized.
After performing: 1) the  transformation to the  relative  and  center of mass (c.m.)
momenta, $\pb$ and $\Pb$, and angular momenta ${\bf l}$ and ${\bf L}$,
and 2) the angular momentum couplings: ${\bf l}+{\bf L}={\mbl}$,
${\mbl}+{\bf S}={\bf J}$ one obtains:
\br
\Gamma_{t_N}&=&\frac{16M_N^3}{\pi}\hat{J}_I^{-2}
\sum_{S\lambda lLTJJ_F{\alpha}}
\int_0^{\Delta_F^{\alpha}}d\e\sqrt{\e(\Delta^{\alpha}_F-\e)}
\nn\\
&\x&\left|\sum_{j_Nj_\Lambda}\M(pPlL\lambda SJT;{j_\Lambda j_N,{t_N}})
\Bra{J_I}\left( a_{j_Nt_N}^\dag a_{j_\Lambda t_\Lambda}^\dag\right)_{J}
\Ket{J_F^{\alpha}}\right|^2,
\label{2.2}\er
where  $\hat{J}\equiv\sqrt{2J+1}$,
$P=2\sqrt{M_N\e}$, $p=\sqrt{M_N(\Delta-\e)}$,  $\Delta^{\alpha}_F=\E_I
-\E^{\alpha}_F-2M_N$,  $t_\Lambda=-1/2$, $t_p=1/2$, $t_n=-1/2$, and the label $\alpha$ goes
over all final states with the same spin and parity. Thus, that the NMD rates, in principle,
depend on both: i) the nuclear structure effects
through the two-particle $N\Lambda$ parentage coefficients
$\Bra{J_I}\left( a_{j_N { t_N}}^\dag a_{j_\Lambda t_\Lambda}^\dag\right)_{J}\Ket{J_F}$, and
on the transition potential via the elementary transition amplitudes
\br
\M(pPlL\lambda SJT;{j_\Lambda j_N,{t_N}})&=&\sqi [1-(-)^{l+S+T}]
\nn\\
&\x&({pPlL\lambda SJ;TM_T=t_\Lambda+t_N}|V|{j_\Lambda j_NJ;t_\Lambda t_N}).
 \label{2.3}\er
Here $(\cdots|V|\cdots)$ is the direct matrix element and the factor
in front takes care of the antisymmetrization.
To evaluate the  nuclear matrix element
one has to carry out the $jj-LS$ recoupling and  the Moshinsky
transformation \cite{Mo59} on the ket $|{j_\Lambda j_NJ})$:
\br
|j_\Lambda j_NJ)
&=&\hat{j}_\Lambda\hat{j}_N
\sum_{\lambda' S'{\sf nlNL}}
\hat{\lambda'}\hat{S'} \ninj{l_\Lambda}
{\fot}{j_\Lambda}{l_N}{\fot}{j_N}{\lambda'}{S'}{J}
 ({\sf nlNL}\lambda'|n_\Lambda l_\Lambda n_Nl_N\lambda') |{\sf nlNL}
\lambda' S'J),
\label{2.4}\er
where $(\cdots|\cdots)$ are the Moshinsky brackets \cite{Mo59}, and
${\sf l}$ and ${\sf L}$ stand for  the quantum numbers of
the  relative and  c.m. orbital angular momenta in the
$\Lambda N$ system.
The explicit expressions for the transition potentials are
given in the Ref. \cite{Ba02}.

When the hyperon  is assumed to be weakly coupled to the $A-1$ core,
which implies that the interaction of $\Lambda$ with core nucleons is 
disregarded, one has that $\ket{J_I}\equiv\ket{(J_Cj_\Lambda)J_I}$,
where   $J_C$ is the spin of the core, and gets
\br
\Bra{J_I}\left( a_{j_N { t_N}}^\dag a_{j_\Lambda
  t_\Lambda}^\dag\right)_{J}\Ket{J_F}
&=&(-)^{J_F+J+J_I}\hat{J}\hat{J}_I
\sixj{J_C}{J_I}{j_\Lambda}{J}{j_N}{J_F}
\Bra{J_C}a_{j_N{ t_N}}^\dag\Ket{J_F}.
\label{2.5}\er
To evaluate the one-particle spectroscopic amplitudes
$\Bra{J_C}a_{j_N{ t_N}}^\dag\Ket{J_F}$ we will use the BCS approximation and,
it will be assumed that the even-even, odd-even and odd-odd cores are 
described, respectively, as  zero, one and two quasiparticle
states. Correspondingly,   the state $\ket{J_C}$ goes  into $\ket{BCS}$,
$b^{\dag} _{j_1t_1}\ket{BCS}$ and $(b^{\dag} _{j_1t_1}
b^{\dag} _{j_2t_2})_{J_2}\ket{BCS}$,
where $\ket{BCS}$ is the BCS vacuum and  $b_{j}^\dag=u_j 
a_{j}^\dag-v_ja_{\overline{j}}$ is the quasiparticle creation operator 
\cite{Ba02}.
In all three  cases the NMD rate can be cast in the form:
\br
\Gamma_{{t_N}}&=&\sum_{j_N}\sum_{ J=|j_N-j_\Lambda|} ^{J=j_N+j_\Lambda}v^{2}_{j_N}F_J(j_Nt_N)
\R_J(j_Nt_N),
\label{2.6}\er
where
\br
\R_J(j_Nt_N)=\frac{16M_N^3}{\pi}\int_0^{\Delta_{j_Nt_N}}d\e\sqrt{\e(\Delta_{j_Nt_N}-\e)}
\sum_{SlL\lambda T}
\M^2(pPlL\lambda SJT;{j_\Lambda j_N,{t_N}}),
\label{2.7}\er
and
\be
\Delta_{j_N{ t_N}}=M_\Lambda-M_N+\e_{j_\Lambda t_\Lambda}+\e_{j_N{ t_N}}.
\label{2.8}\ee The geometrical factors $F_J(j_Nt_N)$ are:
\bnu
\item {\em Even-Even core}: $\ket{J_C}\go\ket{BCS}$,
\br
F_J(j_Nt_N)&=&\hat{j}^{-2}_\Lambda
\hat{J}^{2}
\label{2.9}\er
\item {\em Odd-Even core}: $\ket{J_C}\go b^{\dag} _{j_1t_1}\ket{BCS}$,
\br
F_J(j_Nt_N)&=&\hat{J}^2\sum_{ J_F=|j_1-j_N|}^{j_1+j_N}
\left[1+(-)^{J_F}\delta_{j_1j_N}\delta_{t_Nt_1}\right]\hat{J}_F^2
\sixj{j_1}{j_N}{J_F}{J}{J_I}{j_\Lambda}^2.
\label{2.10}\er
\item {\em Odd-Odd core}: $\ket{J_C}\go (b^{\dag} _{j_1t_1}b^{\dag} _{j_2t_2})_{J_2}\ket{BCS}$,
\enu
\br
F_J(j_Nt_N)&=&\delta_{t_N{t_1}}\hat{J}^2\hat{J}^2_2
\sum_{ J_F=|J_1-j_2|}^{J_1+j_2}\hat{J}^2_F\sum_{
  J_1=|j_1-j_N|}^{j_1+j_N}
\hat{J}_1^2\left[1+(-)^{J_1}\delta_{j_{1}j_N}\delta_{t_Nt_1}\right]
\nn\\
&&\sixj{J_F}{j_2}{J_1}{j_1}{j_N}{J_2}^{2}
\sixj{J_F}{J_I}{J}{j_\Lambda}{j_N}{J_2}^{2}+(j_1t_1)\leftrightarrow (j_2t_2)
\label{2.11}\er
It is clear that in \rf{2.11} $t_2=-t_1$.

\begin{table}[h]
\begin{center}
\caption{Values of $(2j_N+1)F_J(j_Nt_N)$. The quantum numbers  $(j_1,t_1,J_I)$
in odd-even core nuclei are: $(0s_{1/2},-1/2,0)$ for $^{4}_\Lambda He$,
$(0s_{1/2},1/2,0)$ for $^{4}_\Lambda H$, $(0p_{3/2},-1/2,1)$ for 
$^{12}_\Lambda C$, $(0p_{1/2},-1/2,1)$ for $^{16}_\Lambda O$, and 
$(0d_{5/2},-1/2,2)$ for $^{28}_\Lambda Si$.
Similarly,  $(j_1,t_1,j_1,t_1;J_2J_I)$ in odd-odd core nuclei are: 
$(0s_{1/2},-1/2,0s_{1/2},1/2;1,1/2)$ for $^{3}_\Lambda He$, and 
$(0p_{3/2},-1/2,0p_{3/2},1/2;3,5/2)$ for $^{11}_\Lambda B$.}
\vspace{1cm}
\begin{tabular}{|ccc|ccccccccc|}
\hline
$j_N$   & $t_N$ &$J$&$^{3}_\Lambda H$ &$^{4}_\Lambda He$&$^{4}_\Lambda H$
&$^{5}_\Lambda He$&$^{11}_\Lambda B$&$^{12}_\Lambda C$&$^{16}_\Lambda
  O$&$^{17}_\Lambda O$&$^{28}_\Lambda Si$\\
\hline
  $0s_{1/2}$&n  &$0$ &$3/2$ &$2$ &$1$ &$1$ &$1$ &$1$&$1$&$1$&$1$\\
\hline
            &n  &$1$ &$1/2$ &$0$ &$3$ &$3$ &$3$  &$3$&$3$&$3$&$3$\\ \hline
            &p  &$0$ &$3/2$ &$1$ &$2$ &$1$ &$1$   &$1$&$1$&$1$&$1$\\ \hline
            &p  &$1$ &$1/2$ &$3$ &$0$ &$3$ &$3$   &$3$&$3$&$3$&$3$\\ \hline
\hline
$0p_{3/2}$  &n  &$1$ &$-$ &$-$ &$-$   &$-$ &$13/2$&$7$ &$6$&$6$&$6$\\
\hline
 & n & $2$&$-$&$-$&$-$&$-$&$11/2$&$5$&$10$&$10$&$10$\\ \hline
 & p& $1$&$-$&$-$&$-$&$-$&$13/2$&$6$&$6$&$6$&$6$\\ \hline
& p& $2$&$-$&$-$&$-$&$-$&$11/2$&$10$&$10$&$10$&$10$\\ \hline
\hline
$0p_{1/2}$  &n  &$0$ &$-$ &$-$ &$-$   &$-$ &$-$&$-$ &$0$&$1$&$1$\\
\hline
 & n & $1$&$-$&$-$&$-$&$-$&$-$&$-$&$2$&$3$&$3$\\ \hline
 & p& $0$&$-$&$-$&$-$&$-$&$-$&$-$&$1$&$1$&$1$\\ \hline
& p& $1$&$-$&$-$&$-$&$-$&$-$&$-$&$3$&$3$&$3$\\ \hline
\hline
$1d_{5/2}$  &n  &$2$ &$-$ &$-$ &$-$   &$-$ &$-$&$-$ &$-$&$-$&$16$\\
\hline
 & n & $3$&$-$&$-$&$-$&$-$&$-$&$-$&$-$&$-$&$14$\\ \hline
 & p& $2$&$-$&$-$&$-$&$-$&$-$&$-$&$-$&$-$&$15$\\ \hline
& p& $3$&$-$&$-$&$-$&$-$&$-$&$-$&$-$&$-$&$21$\\
\hline\end{tabular}
\end{center}
\label{table1}
\end{table}
The  nuclear structure
manifests  itself basically through the factors $F_J(j_Nt_N)$, which are
engendered by the Pauli principle. Their values for a few cases
are given in Table I. We would like to stress that the quantum numbers
${j_1t_1}$ and ${j_2t_2}$ stand for the hyperon partners in the
initial state, and that $j_N$ runs over all proton and neutron
occupied 
states in the initial nucleus. It is amazing to notice that the the
last three equations are  valid for any hypernucleus,
which could be so light as   ${^{3}_\Lambda H}$ or so heavy as
${^{208}_\Lambda Pb}$.  One should also add that the Eq. \rf{2.6}
contains the same physics as the Eq. (5) in Ref. \cite{Pa97} or the
Eq. (30) in Ref. \cite{It02}, with the advantage that we do not have
to deal with  spectroscopic
factors. Of course, neither the initial and final wave functions are
needed.  From the results displayed in  Table I
it can be seen that in all three cases the coefficients $F_J(j_Nt_N)$
are of the same order of magnitude, which indicates that the nuclear
structure effects in the NMD are of minor importance.
\newpage
\section{Numerical Results and Discussion}
The numerical values of the parameters  necessary to specify the
transition potential, were taken  from Ref. \cite{Pa97}, where, in
turn, the strong couplings have been taken from Refs. \cite{Na77,Ma89}
and the  weak ones from Ref. \cite{Du96}.
The energy difference $\Delta_{j_N{t_N}}$ in \rf{2.8} is evaluated
from the 
experimental
single nucleon and hyperon energies.
It is a general belief nowadays that, in any realistic evaluation of the
 hypernuclear NMD rates, the  finite nucleon size (FNS) and the
  short range correlations (SRC) have to be included simultaneously.
 Therefore, in the present paper both the FNS and SRC
 renormalization effects are considered,  in the way described in
 Ref. \cite{Ba02}.
Under these circumstances, and because of the relative smallness of pion mass,
 the transition is dominated by the OPE  \cite{Pa97,Ba02}.

The  numerical calculations were done in the extreme shell model, which implies
that the pairing factors $v_{j_N}$ were taken to be equal to one (zero) for the
occupied (empty) levels. Thus, from the nuclear structure point of view, the
only  free parameter  is  the harmonic oscillator length $b$. We
evaluate it from the relation $b=1/\sqrt{\hbar \omega M_N}$, and the 
oscillator  energy was estimated from the relation 
$\hbar\omega=45A^{-1/3}-25A^{-2/3}$ MeV, which is frequently used for
light  nuclei.
\begin{table}[h]
\begin{center}
\caption{
Nonmesonic decay rates $\Gamma_{NM}$ in units of $\Gamma^0= 2.50 
\cdot 10^{-6}$ eV. The symbols $\pi$, PS and V stand, respectively,
for the transition potentials activated by the pion, the pseudo-scalar
($\pi+\eta+K$), and  the vector ($\rho+\omega+K^*$) mesons.
Experimental data are shown for comparison.
\label{table2}}
\vspace{1cm}
\begin{tabular}{|c|cccc|}\hline
Hypernucleus&$\pi$&$PS$
&$PS+V$&$EXP$\\ \hline
$^{3}_\Lambda H$ &$0.154     $ &$ 0.107  $&$0.140$&$$\\
$^{4}_\Lambda He$ &$0.546          $ &$0.357      $&$0.507   $&$$\\
$^{4}_\Lambda H$ &$ 0.106         $ &$0.192    $&$0.168    $&$$\\
$^{5}_\Lambda He$ &$0.553              $ &$ 0.508        $&$0.609    $
&$0.41\pm 0.14$~\cite{Sz91}\\
$^{11}_\Lambda B$ &$0.835     $ &$ 0.737       $&$ 0.880        $
&$0.95\pm 0.13\pm 0.04$\cite{No95}\\
$^{12}_\Lambda C$ &$0.971       $ &$   0.820        $&$ 1.000            $
 &$1.14\pm 0.2$\cite{Sz91}\\
&&&&$0.89\pm 0.15\pm 0.03$~\cite{No95}\\
&&&&$1.14\pm 0.08$~\cite{Bh98}\\
$^{16}_\Lambda O$ &$ 1.136  $ &$   0.969          $&$ 1.171

$&$$\\$^{17}_\Lambda O$ &$1.178       $ &$  1.028         $&$ 1.226

          $&$$\\
$^{28}_\Lambda Si$ &$1.314          $ &$  1.100      $&$  1.322         $
&$1.30\pm0.10$~\cite{Pa00}\\\hline
\end{tabular}
\end{center}
\end{table}
The calculations for $\Gamma_{NM}$ and $\Gamma_{n/p}$ are
confronted the  experimental data in Tables \ref{table2} and in \ref{table3},
respectively.
Three different OMEM have been employed for the transition potential. Namely:
($\pi$) only the pion was taken into the account, ($PS$) all three 
pseudoscalar mesons ($\pi+\eta+K$) were included, and  ($PS+V$) also
the vector ($\rho+\omega+K^*$) mesons are considered.
The same remarks are pertinent here as in the study \cite{Ba02} where only  
$^{12}_\Lambda C$ has been analyzed. That is:
(1) the simple OPE model accounts  for $\Gamma_{NM}$, but it fails badly
regarding $\Gamma_{n/p}$,
(2) when   $\eta$ and $K$ mesons are included, the
total transition rate is  only slightly modified, while $\Gamma_{n/p}$
change significantly, coming somewhat closer to the measured values,
and (3) the results are not drastically modified when all vector
mesons are built-in.
\begin{table}[ht]
\begin{center}
\caption{
The $p/n$ ratios for several hypernuclei. See the Table II caption.
\label{table3}}
\vspace{1cm}
\begin{tabular}{|c|cccc|}
\hline Hypernucleus&$\pi$&$PS$
&$PS+V$&$EXP$\\ \hline
$^{3}_\Lambda H$ &$    0.491$ &$   0.664 $&$0.506$&$$\\
$^{4}_\Lambda H$ &$   2.996$ &$  24.60  $&$    10.43  $&$$\\
$^{4}_\Lambda He$ &$ 0.108$ &$   0.045   $&$  0.061 $&$0.25\pm0.13$~
\cite{Ze98}\\
$^{5}_\Lambda He$ &$0.160 $ &$      0.539   $&$   0.320
$&$0.93\pm0.55$~
\cite{Sz91}\\
&&&&$1.97\pm0.67$~\cite{No95}\\
$^{11}_\Lambda B$ &$ 0.167  $ &$      0.515    $&$        0.318$&
$1.04 ^{+0.59}_{-0.48}$~\cite{Sz91}\\
&&&&$2.16\pm0.58 ^{+0.45}_{-0.95}$~\cite{No95}\\
$^{12}_\Lambda C$ &$   0.137$ &$         0.416      $&$      0.258   $&
$1.33 ^{+1.12}_{-0.81}$~\cite{Sz91}\\
&&&&$1.87\pm0.59 ^{+0.32}_{-1.00}$~\cite{No95}\\
&&&&$1.17^{+0.09+0.20}_{-0.08-0.18}$~\cite{Ha02}\\
$^{16}_\Lambda O$ &$     0.138$ &$           0.458     $&$
0.279    
$&$$\\
$^{17}_\Lambda O$ &$     0.159$ &$           0.518    $&$        0.315   $&$$\\
$^{28}_\Lambda Si$ &$  0.145$ &$           0.477      $&$          0.294    $&$
1.38^{+0.13+0.27}_{-0.11-0.25}$~\cite{Ha02}\\\hline
 \end{tabular}
\end{center}
\end{table}

\section{Summary and Conclusions}

The shell model formalism for the nonmesonic weak decay of the
hypernuclei involves a partial wave expansion of the emitted nucleon waves
and  preserves   naturally  the antisymmetrization between the
escaping particles and the residual  core. The general expression \rf{2.2} is valid
for any nuclear model and it shows that the nonmesonic transition rates should depend,
in principle, on both: (i) the weak transition potential, through the
 elementary transition amplitudes $\M(pPlL\lambda SJT;{j_\Lambda j_N,{t_N}})$,
 and (ii) the nuclear structure,
through the two-particle $N\Lambda$ parentage coefficients
$\Bra{J_I}\left( a_{j_N { t_N}}^\dag a_{j_\Lambda }^\dag\right)_{J}\Ket{J_F}$.
The latter  explicitly depend on the initial and final
 wave functions. Yet, as explained in Ref. \cite{Ba02}, and because of:
 a) the inclusive nature of the nonmesonic decay, and
 b) the peculiar properties of the coefficients $F_J(j_Nt_N)$,
this   dependence is washed out.
In this way we have arrived at a very simple  result for
transition rates, given by the Eq. \rf{2.6}, which is valid, not only for the
hypernuclei with odd-even core  (as shown previously \cite{Ba02}),
but also for those which have even-even and odd-odd cores, which has been demonstrated
here.

We  reproduce satisfactorily  the data for  the total
transition rates with the OMEM parametrization from the literature \cite{Pa97},
but the $n/p$-ratios  are not well accounted for.
Thus, after
having acquired full control of the nuclear structure involved in the process,
and after having convinced ourselves that the  nuclear structure  correlations
can not play a crucial role, we firmly believe that the currently used
OMEM should be radically changed. Either  its parametrization has to   
be  modified or  additional degrees of
freedom have to be  incorporated, such as  the   factorizable terms 
\cite{Fis73}, the axial-vector-meson 
exchanges \cite{Bir96}, or the correlated  the correlated $2\pi$  
from Ref.  \cite{It02}.

\bigskip
The authors acknowledge  the support of   ANPCyT (Argentina) under
grant BID 1201/OC-AR (PICT 03-04296), of CONICET under grant PIP 463,
 of Fundaci\'on Antorchas
(Argentina) under grant  13740/01-111, and of Croatian Ministry of
Science and Technology under grant $\#$ 119222. D.T. thanks the Abdus
Salam ICTP
Visiting Scholar programme for  travel fares. F.K. is
a Fellow of the CONICET  Argentina. We would like to thank
A.P. Gale\~ao, C. Barbero and E. Bauer for very helpful and
illuminating  discussions.


\begin{thebibliography}{99}
\bibitem{Us99} Q.N. Usmani and A.R. Bormer, Phys. Rev. {\bf C60}, 055215 (1999).
\bibitem{Mo74} A. Montwill {\sl et al.}, Nucl. Phys.  {\bf A234}, 413 (1974).
\bibitem{Ni76} K.J. Nield {\sl et al.}, Phys. Rev. {\bf C13}, 1263 (1976).
\bibitem{Gr85} R. Grace  {\sl et al.}, Phys. Rev. Lett. {\bf 55}, 1055 (1985).
\bibitem{Bo87} J.P. Bocquet {\sl et al.}, Phys.  Lett. {\bf B192}, 312 (1987).
\bibitem{Sz91} J. J. Szymanski {\sl et al.}, Phys. Rev. {\bf C43}, 849 (1991).
\bibitem{No95} H. Noumi {\sl et al.}, Phys. Rev. {\bf C52}, 2936 (1995).
\bibitem{Ze98} V.J.Zeps,  Nucl. Phys. {\bf A639} 261c (1998).
\bibitem{Bh98} H. Bhang {\sl et al.}, Phys. Rev. Lett. {\bf 81}, 4321 (1998).
\bibitem{Pa00} H. Park {\sl et al.}, Phys. Rev. {\bf C61}, 054004 (2000).
\bibitem{Ha02} O. Hashimoto  {\sl et al.}, Phys. Rev. Lett. {\bf 88}, 042503 (2002).
\bibitem{Jun02} J-H. Jun, Phys. Rev. {\bf C63}, 044012 (2001);
\bibitem{Ad67} J.B. Adams, Phys. Rev. {\bf 156}, 1611 (1967).
\bibitem{Mc84} B. H. J. McKellar and B. F. Gibson, Phys. Rev. {\bf
  C30}, 322 (1984).
\bibitem{Ta85} K. Takeuchi, H. Takaki and H. Band$\bar{\rm o}$,
Prog. Theor. Phys. {\bf 73} (1985) 841.
\bibitem{He86} D.P. Heddle and L.S. Kisslinger, Phys. Rev. {\bf C33}, 608 (1986).
\bibitem{Ba90} H. Band\~o, T. Motoba and \v Zofka,  Int. J. Mod. Phys. {\bf A21} (1990) 4021.
\bibitem{Co90} J. Cohen, Prog. Part. Nucl. Phys. {\bf 25}, 139, edited by
A. Faessler, (Pergamon, 1990).
\bibitem{Al91} W.M. Alberico, A. De Pace, M. Ericson and A. Molinari,
Phys. Lett. {\bf B256}, 134 (1991).
\bibitem{Ra92} A. Ramos, E. van Meijgaard, C. Bennhold and B.K. Jennings, Nucl. Phys.
{\bf A644}, 703 (1992).
\bibitem{Ra95} A. Ramos, E. Oset, and L. L. Salcedo,
Phys. Rev.  {\bf C50}, 2314 (1995).
\bibitem{Pa95} A. Parre\~{n}o, A. Ramos and E. Oset,
Phys. Rev. {\bf C51}, 2477 (1995).
\bibitem{Pa95a} A. Parre\~{n}o, A. Ramos and C. Bennhold, Phys. Rev. {\bf C52},
R1768 (1995): {\bf C54}, 1500 (E) (1996).
\bibitem{Du96} J. F. Dubach, G. B. Feldman, B. R. Holstein and L. de la Torre,
Ann. Phys. (N.Y.) {\bf 249}, 146 (1996).
\bibitem{Pa97} A. Parre\~{n}o, A. Ramos and C. Bennhold,
Phys. Rev. {\bf C56}, 339 (1997).
\bibitem{Ra97} A. Ramos, M.J. Vicente-Vacas and E. Oset,
Phys. Rev.  {\bf C55}, 735 (1997).
\bibitem{Os98} E. Oset and A. Ramos, Prog. Part. Nucl. Phys. {\bf 41}, 191, edited by
A. Faessler, (Pergamon, 1998).
\bibitem{Pa98} A. Parre\~no, A. Ramos, C. Bennhold and K. Maltman, Phys. Lett.
{\bf B 435}, 1 (1998).
\bibitem{In98} T. Inoue, M. Oka, T. Motoba and K. Itonaga, Nucl. Phys.
{\bf A633}, 312 (1998).
\bibitem{It98} K. Itonaga, T. Ueda, T. Motoba, Nucl. Phys.
{\bf A639}, 329c (1998).
\bibitem{Pa99} A. Parre\~{n}o, A. Ramos, N.G. Kelkar and C. Bennhold,
Phys. Rev. {\bf C59}, 2122 (1999).
\bibitem{Alb00} W. M. Alberico and G. Garbarino,
Phys. Lett. {\bf B486}, 362 (2000).
\bibitem{Al00} W. M. Alberico,  A. De Pace, G. Garbarino, and A. Ramos,
Phys. Rev. {\bf C61}, 044314 (2000).
\bibitem{Sa00} K. Sasaki, T. Inoue and M. Oka, Nucl. Phys.
{\bf A669}, 331 (2000), Erratum-ibid {\bf A678}, 455 (2000).
\bibitem{Sa00a} K. Sasaki, T. Inoue and M. Oka, Nucl. Phys. {\bf A678}, 455 (2000).
\bibitem{Pa01} A. Parre\~no and A. Ramos, Phys. Rev. {\bf C65}, 015204 (2001);
A. Parre\~no, A. Ramos and C. Bennhold, Phys. Rev. {\bf C65}, 015205
 (2001)
\bibitem{Os01} E. Oset, D. Jido and J.E. Palomar,
 Nucl.Phys. {\bf A691}, 146 (2001); D. Jido, E. Oset and J.E. Palomar,
arXiv nucl-th/0101051.
\bibitem{It02} K. Itonaga, T. Ueda, T. Motoba, Phys. Rev. {\bf C65}, 034617 (2002).
\bibitem{Ba02} C. Barbero, D. Horvat, F. Krmpoti\'c, T. T. S. Kuo, Z. Naran\v ci\'c and
D. Tadi\'c, Phys. Rev.  {\bf C66}, 055209 (2002).
\bibitem{Na77} M.N. Nagels, T.A. Rijken, and J.J. de Swart, Phys. Rev.
{\bf D15}, 2547 (1977).
\bibitem{Ma89} P.M.M. Maessen, Th. A. Rijken and J.J. de Swart,
Phys. Rev.  {\bf C40}, 2226 (1989).
\bibitem{Ha93}D. Halderson, Phys. Rev.  {\bf C48}, 581 (1993).
\bibitem{Pa94}
A. Parre\~no, A. Ramos, C. Bennhold, and D. Halderson, in
{\it Dynamical Features of Nuclei and Finite Fermi
Systems},(World Scientific, Singapore, 1994) p. 318.
\bibitem{De80} B. Desplanques, J. Donoghue and B. R. Holstein,
Ann. Phys. (N.Y.) {\bf 124}, 449 (1980).
\bibitem{Ma69} R. E. Marshak,  Riazuddin and C.P. Ryan:
{\it Theory of Weak Interactions in
Particle Physics}, (Wiley Interscience, New York, 1969).
\bibitem{To82} L. de la Torre, Ph.D. thesis, University of Masschusetts, 1982.
\bibitem{Ok82} L.B. Okun: {\it Leptons and Quarks} (North Holland, 
Amsterdam,1982).
\bibitem{Co83} E.D. Commins and P.H. Bucksbaum: {\it Weak Interactions
  of Leptons and Quarks} (Cambridge University Press, Cambridge, 1983).
\bibitem{Na88} G. Nardulli, Phys. Rev. {\bf C38}, 832 (1988).
\bibitem{Ba01} C. Barbero, D. Horvat, F. Krmpoti\'c, Z. Naran\v ci\' c
  and D. Tadi\'c, Fizika {\bf B10} (2001) 1, and
Fizika {\bf B10} (2001) 307.
\bibitem{Ba01a} C. Barbero, D. Horvat, F. Krmpoti\'c, Z. Naran\v ci\' c, M.D.
Scadron and D. Tadi\'c,\\
 Jour. Phys. G {\bf 27} (2001) B21.
\bibitem{Br02} G.E. Brown and M. Rho, Phys.Rept. {\bf 363}, 85 (2002).
\bibitem{Mo59} M. Moshinsky,  Nucl. Phys. {\bf 13} 104 (1959).
\bibitem{Fis73} E. Fischbach and D.Tadi\'c Phys. Rep. {\bf C6} 125 (1973).
\bibitem{Bir96}   M. Birkel and H. Fritzsch Phys. Rev. {\bf D53} 6195 (1996).
\end{thebibliography}
\end{document}